# INTELLIGENT INTERNET OF THINGS (IOT) NODE DEMONSTRATOR FOR DEVICE MONITORING AND CONTROL IN THE OIL AND GAS SECTOR

Stephen Ugwuanyi
and
James Irvine

**ABSTRACT**

*Internet of Things (IoT) is the new industrial slogan for connecting intelligent and unintelligent devices to the web. The problem of security of data transfer, interoperability of different proposed methodologies, the ubiquity of Wi-Fi and the development of low power consuming MCUs has broadened the search for the best alternative technology for IoT in the oil and gas sector. This paper focus on the communication method for IoT devices to determine the level of functionality and the efficiency of interfacing the new MOD-WIFI-ESP8266-DEV Wi-Fi unit based on the IEEE 802.11 standard with MSP430 by Texas Instrument. The system controls LEDs and monitors Temperature/Humidity sensor (DHT11) using Android application and web service. The system presents in three-layered structure an ecosystem of lightweight, small size, reduced cost and low power IoT system. It is expected that industries/users of this system would be able to control, monitor, and analyse data generated by the web of connected devices.*

## Introduction

Internet of Things (IoT) according to International Telecommunication Union (ITU) and International Energy Research Centre (IERC) in (Abdul-Rahman and Graves, 2016) is, "a dynamic global network infrastructure with self-configuring capabilities based on standard and interoperable communication protocols where physical and virtual "things" have identities, physical attributes, and virtual personalities, use intelligent interfaces and are seamlessly integrated into the information network". It gives objects the capabilities to auto-configure themselves, communicate and carry out data processing. The number of these heterogeneous devices ranging from sensors nodes to appliances in home and offices, smartphones expected to be linked by various technologies such as wireless networks is now on the increase, thus giving birth to IoT (Baccelli, et al., 2013; Hutchison and Mitchell, 2008).

The majority of these heterogeneous devices added to the internet are from different companies with their concept of IoT. The means of connecting these devices to the internet and having them seamlessly communicate with each other without a human assistant is no longer an impossible task to achieve. The challenge of building a common communication software and application to handle the existing unrelated software and modules is now the biggest problem in IoT (Hutchison and Mitchell, 2008).

The development of the internet over the years is based on Internet Protocol Version 4 (IPv4) which is now being replaced by Internet Protocol Version 6 (IPv6), making every device on the web to have its Internet Protocol (IP) address (Steurer, 2010; Gershenfeld, et al., 2004; Ugwuanyi, et al., 2017). The increased address space of IPv6 gave rise to Machine-to-Machine (M2M) communication, enabling devices to talk to one another and act upon information autonomously. This brought a drastic change both within the scope and scale of the internet. According to (Lab, 2012), industry leaders predicted that the number of connected devices would surpass 15 billion nodes by 2015 and reach over 50 billion by 2020, see Fig.1.

Consequently, this posed a future challenge for the players within the embedded system industry on how to unlock the potentials of this growing interconnected web of devices, known as IoT. According to Metcalfe's Law as reported by (Lab, 2012), the value of a network is equal to the square of the number of devices connected to it. A system design and implementation considering selected factors with regards to wireless connectivity is therefore necessitated.

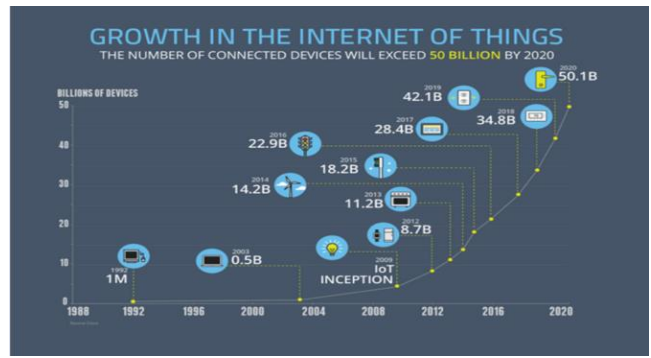

Fig. 1. IoT growth [National Cable Telecommunication Association 2016]

## IOT REVIEW

From the conceptual review of the methodologies, it is seen that IoT has different means of communication through which a node can be connected to the internet. The reviewed literature reveals that many researches have been implemented using wired, sensor and actuators network, and different wireless technologies interfaced with various MCUs to form a complete ecosystem of IoT. Ahmed and Corey presented an IoT system that successfully used Bluetooth and cloud application, ThingSpeak, a cloud-based application to remotely monitor the operation of the networked devices through a compauter-based application, C-sharp programming language. Ronan in (Ronan, 2015), used wired technology and Bluetooth to achieve the same result. In (Anis, et al., 2016; Thomas, et al., 2016), Wi-Fi technology but different MCU and methodology were used to achieve the same result.

Most importantly, it is seen that all the reviewed literatures identified the need for an improved wireless technology for the IoT technology. Wi-Fi is considered to be the most widely recommended technology expected to fast track its rapid development considering the level of deployment of such network; it advantages when compared to other wireless technologies. The only challenge as mentioned in (Lab, 2012) is an unexpected increase in the power consumption as the numbers of devices linked to the web increases. The work of (Gonchigsumlaa, et al., 2015), shows that an improvement of Wi-Fi energy consumption is underway.

Consequently, the project implementation adopted the method similar to (Ronan, 2015; Thomas et al., 2016; Abdul-Rahman & Graves, 2016; Lab, 2012). Achieved by interfacing MOD-WIFI-ESP8266-DEV Wi-Fi unit to the MSP430G2553 control unit. ESP8266 is based on the IEEE 802.11 standards seen as a potential technology for M2M communication. Analysing various technologies through which a wireless interface can be added to the control unit having sensors and LED as end nodes. The monitoring of the sensor and the control of the LEDs is achieved through a web server interface implemented using active web server and an Android Application built using Android Studio.

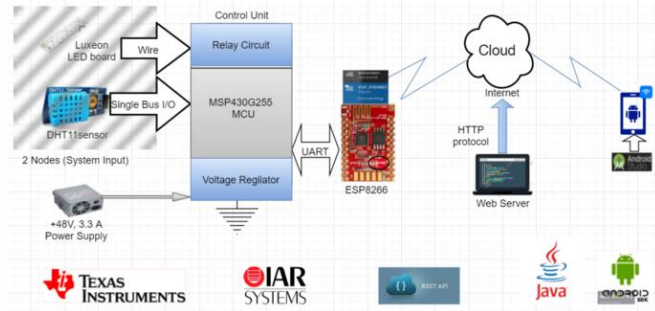

**Fig. 2: System Block Diagram Showing the Major Components**

The system diagram forms an ecosystem core comprising of the hardware devices and gateways and software platforms as well as the standard protocol for enabling interconnections. This will make achievable the requirement for an intelligent node; being always aware of the node state and the possibilities of taking appropriate feedback control actions.

**Internet of Things (IoT) in the Oil and Sector**

Companies in oil and gas sector are constantly faced with problems that severely impact their overall value chain. Internet of Things is a new solution now widely adopted to improve oil exploration, maintain existing facilities, reduce environmental pollution, and most importantly to reduce nonproductive time. The introduction of IoT to one Shell Nigeria oilfield contributed more than $1 million to their business in 2017 ("Shell Boosts Business by Drilling for Data," n.d.). The IoT platform was only introduced to monitor wellheads and pipelines. Billions of dollars can be save by deploying this technology. Real time data processing will aid preventive maintenance, accident prevention.

According to (Goranson, et al., 1997), the protection and management of data generated within an oil field have significant operational advantage. The applications of IoT in the oil and gas sector has no boundary. The upstream, midstream, and downstream will benefit from this technology in the following ways:
- Upstream – optimize process, real-time data processing, and reduce nonproductive time.
- Midstream – effective monitoring of petroleum transporting vehicles, pipeline flowrate monitoring and leak detection.
- Downstream – exploration infrastructural monitoring for proper shutdown, improve safety, and revenue increase.

## MATERIALS AND METHODS

**System Design**

MSP40 control unit according to (Ronan Foubert, 2015) shown in Fig. 3 is the starting point for the project implementation. It is made up of voltage regulators to control the amount of voltage supplied to the MCU. TMR 3-4811 is a DC-DC 3W converter that reduces the 48V input to 5VDC and the current to 600mA. IE0505S is the second DC-DC converter that finally give out a regulated 3.3V; 300mA used to power the MCU and other components. An optocoupler 6N137 handles the utility PWM signal amplification of the MCU. The four LDD-1000H are DC-DC constant current driver relay circuit to support the driving of the LEDs.

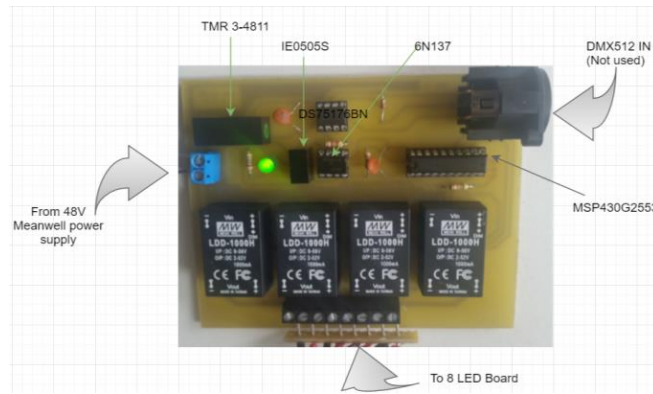

**Fig. 3: MSP430G2553 Control Unit Showing the Input/output Connections**

The Wi-Fi module for the building of this project is the Olimex (not-Expressif vendor) MOD-WIFI-ESP8266-DEV which is based on the ESP-8266 having features of low cost, ease of suitable coding onto the chip and the provision of pin connections. ESP8266 has the features to be set up in such a way that it looks up the readings from the sensor connected to the MSP430 control unit and takes the reading. Also, it functions as a stand-alone controller with the sensor when the coding capacity is within 2MB (16Mb).

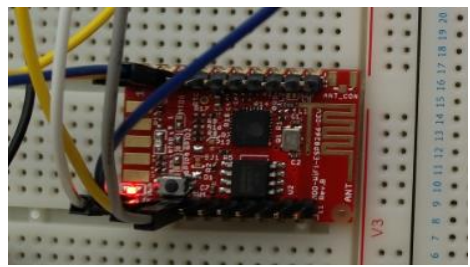

**Fig. 4: A Functional MOD-WIFI-ESP8266-DEV Wireless Module**

**Implementation**

The tested individual components that make up the project system is shown in Fig. 5. The MSP430G2553, MOD-WI-FI-ESP8266-DEV are the processing devices that external nodes were attached to through the UART based serial communication. The devices were fast, and efficient in data transfer at 115200 baud rate.

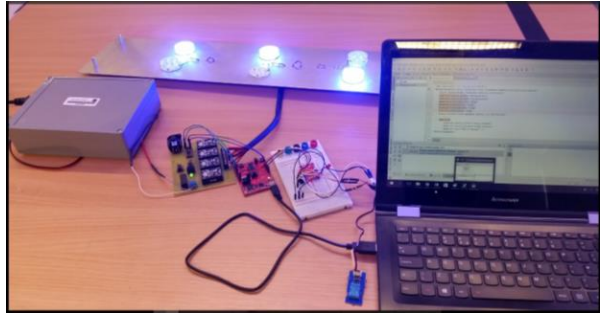

**Fig. 5: Functional Integrated System Hardware Showing all the Devices Used in the Design**

Table 1. below show the pin connections used to setup and reboot the ESP8266.

**Table 1: MOD-WIFI-ESP8266-DEV Setup Pin Configurations**

| ESP8266-DEV | Connection |
|---|---|
| Pin 1 – VCC | 3.3V |
| Pin 2 – GND | GND |
| Pin 3 – GPIO1 | RXD on the MSP430 Launchpad |
| Pin 4 – GPIO3 | TXD on the MSP430 Launchpad |
| Pin 13 – RSTB | Pull to GND to reset. RTS on the MSP430 Launchpad |
| Pin 21 – PIO0 | Pull to GND on boot up to enable FLASH mode. |

Access to the ThingSpeak is made possible using the write Application Peripheral Interface (API) key (W4JX8WHVIQJPNBN9) assigned to the account created and hosted in the ThingSpeak web server.

**RESULTS AND DISCUSSION**

**Remote Monitoring of DHT11**

The remote communication protocol for the entire system is made up of three major steps: Layer that connects the system to the Wi-Fi, layer that connects the Wi-Fi to the internet and layer that allows the interconnected modules talk to the server.

As shown in Fig. 6. ESP8266 broadcast the wireless signal to the environment with access limited by the SSID and password configuration. When a successful connection is established, the user mobile device listens to the appropriate port for packets coming from the Wi-Fi module now in a public Wi-Fi connection and establishes TCP server.

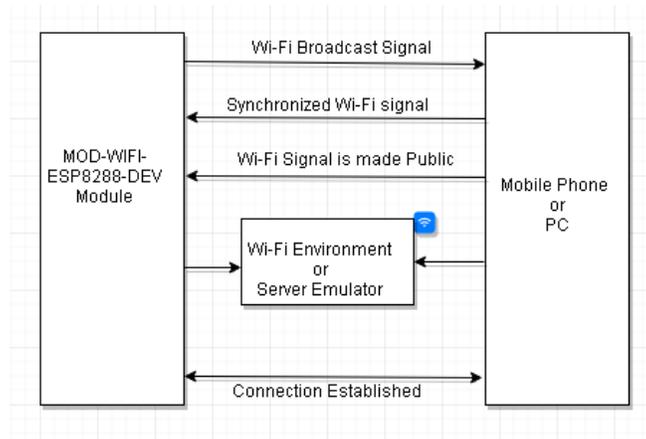

**Fig. 6: Block Diagram of Wi-Fi Connection Establishment**

Within the Wi-Fi environment, the mobile device varies the intensity of the LEDs remotely through TCP port number and IP address directly. When not in the same network environment, the communication is made to pass through the wider network server (cloud) facilitated by the access points depending on the distance between the two nodes. The additional TCP packet overhead and the three-way handshake establishment was regarded as a trade-off for the system stability. The following figures demonstrate the capabilities of remote monitoring of an external sensor node in real time. They show the time interval measures of DHT11 at room temperature in Springburn Glasgow.

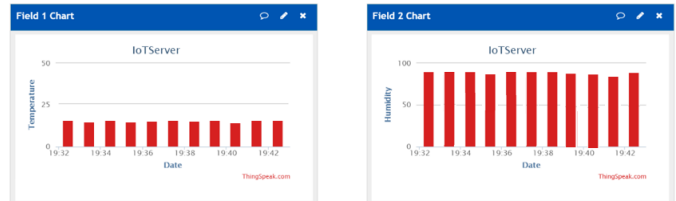

**Fig. 7: Bar Chart Representation of Closely Spaced Temperature and Humidity Data Points**

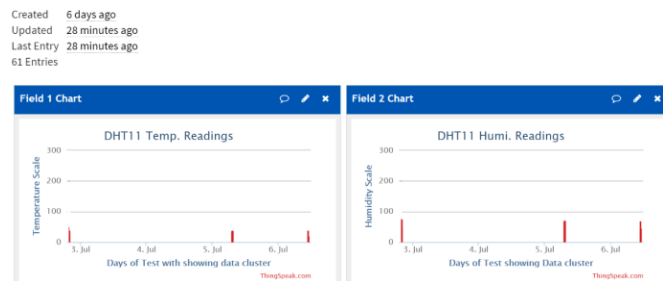

**Fig. 8: 61 Data Points of Temperature and Humidity**

It is observed that both temperature and humidity readings are clustered at a particular value within a given day for better presentation of the data. The experiments show an average temperature of $15^0$C and humidity values of 82% between 8th July 2016 to $14^{th}$ July 2016. The temperature is seen to be relatively stable with a value of 14ºC between $9^{th}$ to $13^{th}$ with 7ºC being the lowest recorded value and 19ºC the maximum value, see Fig. 10.

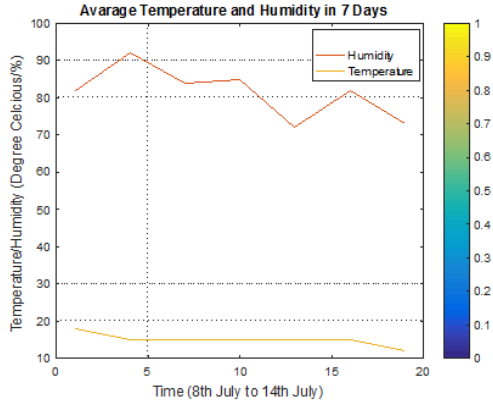

**Fig. 9: Average Temperature and Humidity**

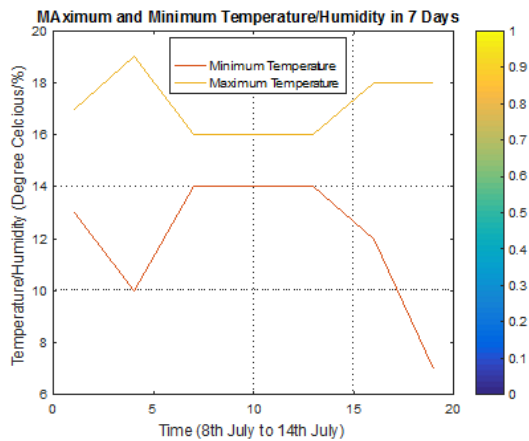

**Fig. 10: Average Maximum and Minimum Temperature**

The average temperature and relative humidity over the duration of the experiment were seen to have a similar shape with the measurement of the United Kingdom online weather company, [www.weatheronline.co.uk].

**Remote Control of LEDs**

The result in Fig. 11. is obtained from the NodeMCU serial monitor. The IP address 192.168.2.1 in Fig. 11 serial monitor, to accessing the interface either on a mobile phone or personal computer that is connected to the internet anywhere in the world. The IP address allocated to the device is dependent on the network access point IP address. The address varies, and it is assigned to the device automatically as soon as it is connected to the network. The control action was observed to be very fast with un-noticed delay. The responsive nature of the whole process is achieved through the aRESP UI library.

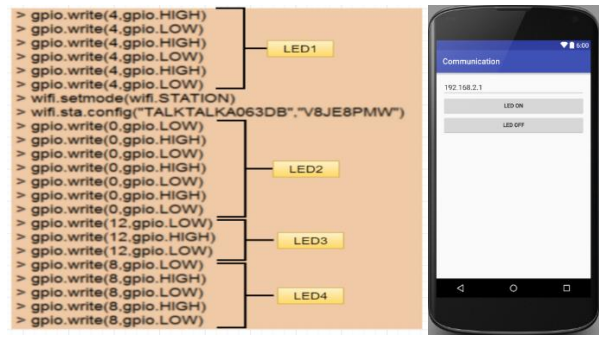

**Fig. 11: NodeMCU Serial Monitor and Android APP Showing the Various GPIOs LEDs Control**

**Conclusion**

IoT is in the centre stage of its revolution that is characterised by heterogeneous nodes and communication standards. According to (Liu & Lu, 2012), it is identified by EU and China as not yet matured but the technology that will bring about industrial and economic development, make life smarter and stimulate the economy.

The proposed IoT node was designed and successfully implemented as the solution for controlling smart devices. The project successfully implements an intelligent IoT node approach to control and monitor sensor nodes in a network on a real time basis from any location through the Wi-Fi approach. This means that is very useful in specific applications like in the the oil and gas and other industrial automation. The project has demonstrated that IoT is capable of taking hold of the major sectors of human development. The goals of this project have been realised, and the results of implemented prototype are satisfactory.

Based on the experimental results, the design model is promising regarding cost and performance. It is probably one of the networking concepts that has the potentials to bring about another industrial IoT revolution with many benefits.

A research sponsored by the Petroleum Technology Development Fund is in progress at the University of Strathclyde, Glasgow to further investigate the security aspect of this solution.